
%
\let\includefigures=\iftrue
%
\let\includefigures=\iffalse
%
\let\useblackboard=\iftrue
%
%
%
\input harvmac.tex
\includefigures
\message{If you do not have epsf.tex (to include figures),}
\message{change the option at the top of the tex file.}
\input epsf
\epsfclipon
\def\fig#1#2{\topinsert\epsffile{#1}\noindent{#2}\endinsert}
\else
\def\fig#1#2{}
\fi
\def\Title#1#2{\rightline{#1}
\ifx\answ\bigans\nopagenumbers\pageno0\vskip1in%
\baselineskip 15pt plus 1pt minus 1pt
\else
\def\listrefs{\footatend\vskip 1in\immediate\closeout\rfile\writestoppt
\baselineskip=14pt\centerline{{\bf References}}\bigskip{\frenchspacing%
\parindent=20pt\escapechar=` \input
refs.tmp\vfill\eject}\nonfrenchspacing}
\pageno1\vskip.8in\fi \centerline{\titlefont #2}\vskip .5in}

\ifx\answ\bigans\def\tcbreak#1{}\else\def\tcbreak#1{\cr&{#1}}\fi
\useblackboard
\message{If you do not have msbm (blackboard bold) fonts,}
\message{change the option at the top of the tex file.}
\font\blackboard=msbm10 scaled \magstep1
\font\blackboards=msbm7
\font\blackboardss=msbm5
\newfam\black
\textfont\black=\blackboard
\scriptfont\black=\blackboards
\scriptscriptfont\black=\blackboardss

\else

\fi
%
\def\yboxit#1#2{\vbox{\hrule height #1 \hbox{\vrule width #1
\vbox{#2}\vrule width #1 }\hrule height #1 }}
\def\fillbox#1{\hbox to #1{\vbox to #1{\vfil}\hfil}}
\def\ybox{{\lower 1.3pt \yboxit{0.4pt}{\fillbox{8pt}}\hskip-0.2pt}}
\def\comments#1{}

\def\p{\partial}
\def\rdel#1{\overleftarrow{\partial_{#1}}}

\def\half{{1\over 2}}
\def\Tr{{\rm Tr\ }}
\def\tr{{\rm tr\ }}

\def\bra#1{{\langle}#1|}
\def\ket#1{|#1\rangle}
\def\vev#1{\langle{#1}\rangle}

\def\Cr{\hfill\break}
\Title{\vbox{\baselineskip12pt
\hfill{\vbox{
\hbox{BROWN-HET-976\hfil}
\hbox{RU-94-89\hfil}
\hbox{hep-th/9412203}}}}}
{\vbox{\centerline{Free Variables and the Two Matrix Model}}}
\centerline{Michael R. Douglas}
\smallskip
\centerline{Dept. of Physics and Astronomy}
\centerline{Rutgers University }
\centerline{\tt mrd@physics.rutgers.edu}
\bigskip
\centerline{Miao Li}
\smallskip
\centerline{Department of Physics}
\centerline{Brown University}
\centerline{Providence, RI 02912}
\centerline{\tt li@het.brown.edu}
\bigskip
\noindent
We study the full set of planar Green's functions for a two-matrix model using
the language of functions of non-commuting variables.
Both the standard Schwinger-Dyson equations and equations determining
connected Green's functions can be efficiently discussed and solved.
This solution determines the master field for the model
in the `$C$-representation.'

\Date{December 1994}
\nref\voi{D.~V.~Voiculescu, K. J. Dykema and A. Nica, {\it Free Random
Variables}, AMS 1992.}
\nref\singer{I.~Singer, ``The Master Field for Two-Dimensional Yang-Mills
theory.''}
\nref\haan{O.~Haan, Z.Physik C6 (1980) 345.}
\nref\cls{P.~Cvitanovic, Phys.Lett. 99B (1981) 49;\Cr
P.~Cvitanovic, P.~G.~Lauwers and P.~N.~Scharbach, Nucl.Phys. B203
(1982) 385.}
\nref\dougrev{M.~R.~Douglas, ``Large N Gauge Theory -- Expansions and
Transitions,'' to appear in the proceedings of the 1994 ICTP Spring School,
hep-th/9409098.}
\nref\mike{M.~R.~Douglas, ``Stochastic Master Fields,'' RU-94-81,
hep-th/9411025.}
\nref\gg{R.~Gopakumar and D.~J.~Gross, ``Mastering the Master Field,''
PUPT-1520, hep-th/9411021.}
\nref\bipz{E. Brezin, C. Itzykson, G. Parisi and J.-B. Zuber,
Comm.Math.Phys 59 (1978) 35.}
\nref\mehta{M.L. Mehta, Comm. Math. Phys. 79 (1981) 327.}
\nref\alf{J.~Alfaro, Phys.Rev.D47 (1993) 4714.}
\nref\matt{M. Staudacher, Phys.Lett.B305 (1993) 332.}
\nref\km{V.~Kazakov and A.~Migdal, Nucl.Phys.B397 (1993) 214.}
%

A recent convergence of work by mathematicians \refs{\voi,\singer}\ and
physicists \refs{\haan,\cls,\dougrev,\mike,\gg} has produced precise
constructions
of the `master field' for general large $N$ gauge and matrix field theories.
The essential difficulty of the large number of degrees of freedom in higher
dimensional large $N$ theories is dealt with by finding master fields which
live in `large' operator algebras such as the type II$_1$ factor associated
with a free group.

When we say that the master field was `constructed' in a higher dimensional
theory, so far what we mean is that if we know all correlation functions for
the theory, we can produce an operator representation of the master field.
Thus to begin we should study models we can already solve, and see if any
simplifications appear in the new language.
The large number of degrees of freedom of higher dimensions (and the need to
use non-trivial free algebras) appears in the large $N$ limit of any integral
over more than one matrix, and the simplest example is the two-matrix model
defined by the large $N$ limit of the integral
\eqn\tmm{Z=\int{\cal D}M_1 \ {\cal D}M_2
 \ e^{-N \Tr[V(M_1) + V(M_2) - c M_1 M_2]}}
with $M_i$ hermitian.  We will take $V$ to be the cubic polynomial $V(M)=\half
M^2 - {g\over 3} M^3$ below, but our considerations generalize easily to higher
order polynomials.  This model was first solved in \mehta\ using orthogonal
polynomials.  This technique uses in an essential way the special form of the
action in \tmm\ and the ability to express the solution in terms of the
invariant observables constructed from a single matrix.  This generalizes to
higher dimensions only in very special cases such as the Kazakov-Migdal model
\km\ and we would like to explore techniques which do not depend so
fundamentally on a particular form for the action.

One such technique is to solve the factorized Schwinger-Dyson equations.
At present, one can do this explicitly only when one can find a subset of these
equations which closes on a few invariants, and such a
reduction for \tmm\ was found in \refs{\alf,\matt}.
We begin by discussing this and the recursive
definition of higher correlation functions in the language of \cls.
The generating function of planar Green's functions is a function of
non-commuting variables
\eqn\defphi{\phi(u_1,u_2) = \sum_{{\rm words}~w} w(u_1,u_2)~
\vev{{1\over N}\Tr w(M_1,M_2)}.}
The Green's functions and thus $\phi$ have cyclic symmetry.\hfil\break
The Schwinger-Dyson equations for \tmm\ then read
\eqn\schdy{\eqalign{\partial_1\phi-c\partial_2\phi-g\partial_1^2\phi=&
\phi u_1\phi\cr
\partial_2\phi-c\partial_1\phi-g\partial_2^2\phi=&
\phi u_2\phi.}}
Derivatives acting on functions of free variables satisfy
\eqn\derivdef{{\p\over\p u_i} u_j f(u) = \p_i u_j f(u) = \delta_{ij} f(u)}
and this is just an alternate notation $\p_i=a_i$, $u_i=a^*_i$ for the free
algebra of \voi. We also define right derivatives
\eqn\rderivdef{f(u) u_j
{\overleftarrow{{\p\over\p u_i}}} = f(u) u_j \rdel{i} = \delta_{ij} f(u).}
%
Acting on a function with cyclic symmetry (such as $\phi$) these satisfy
\eqn\cycprop{\p\ldots f(u) \rdel{i} = \p\ldots \p_i f(u).}

To get a truncated system of equations we work with the first terms in the
series expansion
\eqn\trunc{
\sum_{k\ge 0} u_2^k~\p_2^k\phi(u_1,u_2)\big|_{u_2=0}
=\phi_0(u_1) + u_2\phi_1(u_1)+ u_2^2\phi_2(u_1)+\ldots, }
consider the equations \schdy\ and their derivatives at $u_2=0$,
and look for a minimal set of closed equations.
Identifying terms of the form \trunc\ in \schdy\ we get
\eqnn\schdytrone
\eqnn\schdytrtwo
$$\eqalignno{\partial_1\phi_k-c\phi_{k+1}-g\partial_1^2\phi_k=&
\phi_k u_1\phi_0 &\schdytrone\cr
\phi_1-c\partial_1\phi_0-g\phi_2=& 0.&\schdytrtwo}$$
{}From \schdytrtwo\ a minimal closed set will be
$\{\phi_0,\phi_1,\phi_2\}$
and including \schdytrone\ with $k=0$ and $k=1$ gives three equations.
The operator $\p_1$ is then rewritten
\eqn\minus{\p_1^n f(u_1) = [{1\over u_1^n}f(u_1)]_+
= {1\over u_1^n}\left(f(u_1) - \sum_{i=0}^{n-1} u_1^i~\p_1^if(0)\right)}
producing algebraic equations with boundary conditions $\p_1^i\phi(0)$
for $i\le 3$.
These can be combined into a cubic equation for $\phi_0(u_1)$, to be made more
explicit below.  In general, we get an equation of degree $\deg V$.

This is just a start on the problem of getting all correlation functions and
the simplest description of these would be a single equation for $\phi$ which
`closes.'  Not having very sophisticated techniques for working with functions
of non-commuting variables, what we will mean by this is that we can directly
truncate this single equation to an equation for $\phi_0(u_1)$ and then solve
iteratively for $\phi$ in an expansion with terms containing $u_2$'s in all
possible ways.

We can find such an equation by repeating the strategy of taking derivatives of
\schdy\ and eliminating common factors in the result, but this time not setting
$u_2=0$.  A useful set of equations derived from \schdy\ is
\eqn\hybrid{\partial_2\partial_1\phi-g\partial_2\partial_1^2\phi
-{c\over g}\left(\partial_2\phi-c\partial_1\phi-\phi u_2\phi\right)
=\partial_2\phi u_1\phi,}
and equations derived by acting with the right derivative $\rdel{1}$
\eqn\elimin{\eqalign{\partial_2\phi=&{1\over c}\left(\partial_1\phi-g
\partial_1^2\phi-\phi u_1\phi\right) \cr
\partial_2\partial_1\phi=&{1\over c}\left(\partial_1^2\phi-g
\partial_1^3\phi-\phi u_1\partial_1\phi-\phi\right) \cr
\partial_2\partial_1^2\phi=&{1\over c}\left(\partial_1^3\phi-g
\partial_1^4\phi-\phi u_1\partial_1^2\phi-\partial_1\phi(0)\phi-\partial_1\phi
\right).}}
The final equation is obtained by substituting eqs.\elimin\ into \hybrid:
\eqn\nice{D\phi
=-{c^2\over g}\phi u_2\phi-{c\over g}\phi u_1\phi+\partial_1\phi u_1\phi +
\phi u_1\partial_1\phi-g\partial_1^2\phi u_1\phi-g\phi u_1\partial_1^2\phi-
\phi u_1\phi u_1\phi.}
where the operator $D$ is given by
$$D=-1+g\partial_1\phi(0)+(g-{c\over g}+{c^3\over g})
\partial_1+(1+c)\partial_1^2-2g\partial_1^3+g^2\partial_1^4.$$
First let $u_2=0$ in \nice, to obtain a cubic
equation for $\phi_0(u_1)$. Let $R(z)=1/z \phi_0(1/z)$, this algebraic
equation is re-written as
\eqn\algeb{\eqalign{&F(z,R)\equiv R^3-f(z)R^2+g(z)R-h(z)=0,\cr
f(z)=&-{c\over g}+2z-2gz^2,\cr
g(z)=&g^2z^4-2gz^3+(1+c)z^2-(g+c/g-c^3/g^2)z+1-g\partial_1\phi(0),\cr
h(z)=&g^2z^3+(g^2\partial_1\phi(0)-2g)z^2+(g^2\partial_1^2\phi(0)
-2g\partial_1\phi(0)-1-c)z\cr
&+g^2\partial_1^3\phi(0)-2g\partial_1^2\phi(0)
+(1+c)\partial_1\phi(0)+g-c/g+c^3/g^2.}}
The `initial' data $\partial_1^n\phi(0)$ are not totally independent,
as can be seen from eq.\schdy.
They are determined by requiring the appropriate analytic behavior of
$R(z)$, which for small $g$ must have a single cut on the real axis.
This equation was derived
in \matt\ (see also \alf).

In principle one can solve \nice\ iteratively for $\phi$ in an expansion
\eqn\fullexp{\phi=\sum_{n=0}\phi^{(n)}(u_1,u_2),}
where $\phi^{(n)}$ contains all terms
with $n$ instances of $u_2$.
Since there is only one
term involving $u_2$ explicitly, all terms containing $\phi^{(n)}$
in the $n$-th recursion relation are linear in $\phi^{(n)}$ and depend only
on $\phi_0$.
One can show that this differential operator, linear in $\phi^{(n)}$, has a
unique inverse (under the condition that $\phi^{(n)}$ is cyclic) and thus
$\phi^{(n)}$ is expressed in terms of $\phi^{(m)}$ with $m<n$.

A simpler iterative scheme can be developed by starting with the
(trivial) equation
\eqn\trivones{\phi=1+u_1\p_1\phi+u_2\p_2\phi}
and using \elimin\ to eliminate the terms with $\p_2$ derivatives,
producing
\eqn\trivst{
\phi=1+u_1\p_1\phi+{1\over c}u_2\left(\partial_1\phi-g
\partial_1^2\phi-\phi u_1\phi\right).}
Given $\phi^{(n)}$ as in \fullexp, this equation determines the terms in
$\phi^{(n+1)}$ with an initial $u_2$ (by simply dropping the terms
$1+u_1\p_1\phi$).
There is then a unique cyclically symmetric $\phi^{(n+1)}$ which contains such
terms.

It may be interesting to note that a purely algebraic equation can be derived
for $\phi$.
We start with \trivones\ and its $\p_1$ derivatives
\eqn\trivs{\eqalign{
\phi&=1+u_1\p_1\phi+u_2\p_2\phi\cr
\p_1\phi&=\p_1\phi(0)+u_1\p_1^2\phi+u_2\p_2\p_1\phi\cr
\p_1^2\phi&=\p_1^2\phi(0)+u_1\p_1^3\phi+u_2\p_2\p_1^2\phi.}}
Using \elimin\ to eliminate the terms with $\p_2$ derivatives
produces equations linear in $\p_1^2\phi$, $\p_1^3\phi$ and $\p_1^4\phi$.
One then takes the system of these three equations and \nice, and eliminates
these higher derivatives between the equations, producing a single (non-linear)
equation in terms of $\phi$ and $\p_1\phi$.  One then repeats the same
derivation with the roles of $u_1$ and $u_2$ interchanged, producing an
equation in terms of $\phi$ and $\p_2\phi$.  Finally, these two equations and
\trivones\ can be combined to eliminate $\p_1\phi$ and $\p_2\phi$.  We omit the
final result due to its length.

These manipulations can be carried through despite the non-commuting nature of
$u_1$ and $u_2$.  The final equation however contains the inverses $u_1^{-1}$
and $u_2^{-1}$ which require some discussion.  First, the combinations $u_1
u_2^{-1}u_1$ and $u_2 u_1^{-1}u_2$ appear.  This is incompatible with the
expansion \fullexp\ and such terms must cancel in the final result.  Second,
the
combination $u_2^{-1}\phi$ appears.  It is not clear to us that the iterative
solution \fullexp\ is possible.

It would be interesting to express the results more explicitly.  Of course,
we considered \algeb\ a complete solution of the truncated problem, even though
its solution could be made more explicit, because the techniques for solving
such equations are so familiar.  In practice one rarely even writes down its
exact solution and instead studies limits of it, or its analytic behavior.
Perhaps one should regard the solution of \nice\ in the same spirit.

We now turn to study a similar equation for the master field.
One such equation \mike\ takes the form
\eqn\schwins{{\partial\over \partial M_i} S[\hat{M}]=\eta_i + [A,\hat M_i].}
To deal with this explicitly we must choose a representation for the master
field.
Many representations exist, in principle related by similarity transformations,
and suited for different purposes.
A very natural one is the `$C$-representation,'~\refs{\mike,\gg}
built from $\psi(j)$, the generating functional of connected planar correlation
functions, a function of non-commuting variables $j_i$.~\cls\ %
The master field $\hat{M}_i$ is simply
\eqn\masdef{\hat{M}_i={\p\over\p j_i}+{\p\psi(j)\over\p j_i}}
with the non-commuting derivative as above.
If we define a trace as
\eqn\trdef{\hat \tr \hat O = \hat O ~1\big|_{j=0} = \vev{|\hat O|}}
these operators are a master field.

The relation between the two
generating functionals is \cls
\eqn\relation{\phi(u)=\psi(j), \qquad j_i=u_i\psi(j).}
and an equation with the same content as \schwins\ was derived in \cls\ by
simply changing variables from the Schwinger-Dyson equations.
In general it is
\eqn\schwinss{{\partial\over \partial M_i} S[\hat{M}]\ket{}=j_i\ket{}.}
Comparing with \schwins\ or the similar equation
in \refs{\haan,\gg}, this equation is a bit simpler, as one bypasses the
problem of finding the similarity transformation $A$ (or representing the
algebra $[\hat\Pi_i,\hat M_j]=\delta_{ij}\ket{}\bra{}$).
On the other hand, those equations are representation-independent, while this
one is tied to the $C$-representation.
(The other difference between \schwins\ and the other equations is that it
produces cyclically symmetric equations which sum over all variations of $M_i$
in the original path integral, while the others involve a single variation.)
\hfil\break
For our case \schwinss\ becomes
\eqnn\schdone
\eqnn\schdtwo
$$\eqalignno{\partial_1\psi-c\partial_2\psi-g\partial_1^2\psi
-g(\partial_1\psi)^2=&j_1 &\schdone\cr
\partial_2\psi-c\partial_1\psi-g\partial_2^2\psi
-g(\partial_2\psi)^2=&j_2. &\schdtwo}$$

We first derive an algebraic equation for $\psi_0(j_1)=\psi(j_1,j_2=0)$
by a truncation analogous to \trunc.
Truncating \schdone\ gives
\eqn\numeq{\eqalign{
\partial_1\psi_0-c\psi_1-g\partial_1
(\psi_0\partial_1\psi_0)&=j_1\cr
\partial_1\psi_1-c\psi_2-g\partial_1^2\psi_1
-g\partial_1\psi_1\partial_1\psi_0-g\partial_1\psi(0)\p_1\psi_1
&=0}}
while \schdtwo\ gives
\eqn\numtwo{\psi_1-c\partial_1\psi_0-g\psi_2-g(\psi_1)^2
=0.}
Again we have three equations for three unknowns, and it is easy to see
that a quartic equation can be obtained for $\psi_0$.

In fact this equation
can be directly related to \algeb.
Let $K(j_1)=1/j_1\psi_0(j_1)$
and use the reciprocity relations $j_1=R(z)$, $z=K(j_1)$.
The algebraic
equation $F(z, R(z))=0$ is just $F(K(j_1),j_1)=0$. From \algeb,
it is a quartic equation in terms of $K$.

To solve \schwinss\ more generally, more steps are necessary. Applying
$\partial_2$ to \schdone, then replacing
$\partial_2^2\psi$ by solving \schdtwo, we have
\eqn\hybri{\eqalign{&\partial_2\partial_1\psi-g\partial_2\partial_1^2\psi
-g\partial_2\partial_1\psi\partial_1\psi-g\partial_1\psi(0)
\partial_2\partial_1\psi\cr
&-{c\over g}\left(\partial_2\psi-c\partial_1\psi-j_2\right)+c(\partial_2\psi)^2
=0.}}
One can replace $\partial_2\psi$ in the above equation by solving the first
equation of \schdone. To replace $\partial_2\partial_1\psi$ and
$\partial_2\partial_1^2\psi$, we apply $\partial_1$ to the
first equation from the right, and use the cyclic symmetry of coefficients
of $\psi$. Thus,
\eqn\elimi{\eqalign{\partial_2\psi=&{1\over c}\left(\partial_1\psi-g
\partial_1^2\psi-g(\partial_1\psi)^2-j_1\right)\cr
\partial_2\partial_1\psi=&{1\over c}\left(\partial_1^2\psi-g\partial_1^3\psi
-g\partial_1\psi\partial_1^2\psi-g\partial_1\psi(0)\partial_1^2\psi-1\right)\cr
\partial_2\partial_1^2\psi=&{1\over c}\left(\partial_1^3\psi-g\partial_1^4
\psi-g\partial_1\psi\partial_1^3\psi-g\partial_1^2\psi(0)\partial_1^2\psi
-g\partial_1\psi(0)\partial_1^3\psi\right).}}
Substituting these into \hybri, a single equation containing only derivatives
with respect to $j_1$ is obtained. This equation can be solved
iteratively in the same way as \nice.
Using \masdef, this solution determines the master fields $\hat M_i$ for this
model.  We intend to study these in more detail in subsequent work.

To summarize,
we formulated and solved both large $N$ Schwinger-Dyson
equations and master field equations in the two matrix model,
in terms of functions of non-commuting variables.
The master field equation had the same basic structure as the Schwinger-Dyson
equation and the same truncation method worked to solve it.

It is interesting to ask if truncation can work in more models.
Another model which can be solved by the methods presented here has action
$S(M_1,M_2)=\tr(M_1^2+M_2^2-cM_1M_2M_1M_2)/2$,
and we will discuss this and the general `meanders' problem in future work.

There is no reason to think that similar truncations exist for all interesting
models and it would be very desirable to develop more general techniques to
work with this type of equation.
For physics, it is often better to have qualitative and approximate approaches
to understanding an equation, and if we can precisely define the equations, we
can try to study these directions systematically.  The results presented here
should provide a useful test case.

It is also interesting
to see whether one can go beyond the large $N$ limit with the master field,
for example to incorporate considerations in \ref\antal{A. Jevicki and
J. Rodrigues, Nucl. Phys. B421 (1994) 278, hep-th/9312118.}.

\medskip
\noindent{\bf Acknowledgments}

We would like to thank A. Jevicki,
M. Staudacher and C.-I. Tan for useful discussions.
The research of M.R.D. was supported
by DOE grant DE-FG05-90ER40559, NSF PHY-9157016 and the A.J. Sloan Foundation;
and the research of M.L. was supported by DOE grant DE-FG02-91ER40688-Task A.

\listrefs
\end